\documentclass[sigplan,screen]{acmart}
\usepackage{bo-macros}

\AtBeginDocument{%
  }

\acmYear{2025}
\setcopyright{none}
\acmDOI{}
\acmConference[SAA' 25]{1st Workshop on Systems for Agentic AI}{October 13th}{Seoul, Republic of Korea}

\acmBooktitle{1st Workshop on Systems for Agentic AI (SAA '25), October 13th, 2025, Seoul, Republic of Korea}
\acmISBN{}
\acmDOI{}

\begin{document}

\newcommand{\ours}{\textit{\textsc{Earl}}\xspace}

\title{\texorpdfstring{\ours: \underline{E}fficient \underline{A}gentic \underline{R}einforcement \underline{L}earning \\
Systems for Large Language Models}{\ours: Efficient Agentic Reinforcement Learning Systems for LLMs}}

\author{Zheyue Tan}
\affiliation{%
  \institution{Aalto University}
  \country{}
}
\email{zheyue.tan@aalto.fi}

\author{Mustapha Abdullahi}
\affiliation{%
  \institution{Aalto University}
  \country{}
}
\email{mustapha.abdullahi@aalto.fi}

\author{Tuo Shi}
\affiliation{%
  \institution{Aalto University}
  \country{}
}
\email{tuo.shi@aalto.fi}

\author{Huining Yuan}
\affiliation{%
  \institution{Tsinghua University}
  \country{}
}
\email{yuanhuining0@gmail.com}

\author{Zelai Xu}
\affiliation{%
  \institution{Tsinghua University}
  \country{}
}
\email{zelai.eecs@gmail.com}

\author{Chao Yu}
\affiliation{%
  \institution{Tsinghua University}
  \country{}
}
\email{zoeyuchao@gmail.com}

\author{Boxun Li}
\affiliation{%
  \institution{Infinigence-AI}
  \country{}
}
\email{liboxun@infini-ai.com}

\author{Bo Zhao}
\affiliation{%
  \institution{Aalto University}
  \country{}
}
\email{bo.zhao@aalto.fi}

\begin{abstract}
Reinforcement learning (RL) has become a pivotal component of large language model (LLM) post-training, and agentic RL extends this paradigm to operate as agents through multi-turn interaction and tool use. 
Scaling such systems exposes two practical bottlenecks: 
(1) context length grows rapidly during training, inflating memory usage and latency, and triggering out-of-memory (OOM) failures; 
and (2) intermediate tensors accumulate with context length, making cross-device data movement a major system bottleneck.

We present \textbf{\ours}, a scalable system for efficient agentic RL. \ours designs a \emph{parallelism selector} that dynamically adapts model and training parallelism across RL stages based on sequence length and system load, and a \emph{data dispatcher} that performs layout-aware, decentralized exchange of intermediate data batches. 
Together, these components increase throughput, reduce long-context failures, and enable stable large-scale training of agentic LLMs without relying on hard limits or penalties of context length.
\end{abstract}



\maketitle

\pagestyle{plain}

\section{Introduction}
\label{sec:intro}
Reinforcement Learning (RL) has become a key component in the post-training of large language models (LLMs), used to align model behavior with human preferences~\cite{ouyang_training_2022, christiano2017deep} and to elicit advanced capabilities such as reasoning, tool-use, and decision-making~\cite{guo2025deepseekr1, dong2025reinforcement, team2025kimi}.
Agentic LLMs~\cite{openai_deep_research, gemini_deep_research,team2025kimi,openai_gpt5_2025}, which act as autonomous agents interacting with complex environments, are increasingly prominent and typically trained with agentic RL involving multi-turn interactions and adaptive behavior in response to the environment's feedback, achieving superior reasoning and tool-use performance for real-world applications~\cite{openai_deep_research,gemini_deep_research,jern_agent-q_2025,yu_quasar_2025}.

During RL training, the context length increases dramatically, initially boosting reasoning performance~\cite{guo2025deepseekr1, team2025kimi15, yang_qwen3_2025}, but this introduces significant system-level challenges in memory and communication, limiting overall scalability.
Excessive context growth inflates memory usage and can trigger out-of-memory (OOM) failures.
In agentic RL, this issue is further exacerbated by multi-turn interactions. 
For example, with the Llama-3.1-70B model~\cite{dubey2024llama}, context lengths of 4,096 and 8,196 require around 97~\unit{GB} and 354~\unit{GB} for the training batch, respectively, exceeding the memory capacity of existing GPUs~\cite{ultrascale_playbook}. The memory usage is higher when using KV cache, since it stores additional data for each token.
Existing works typically apply a \textit{hard limit} on maximum context length, and some even introduce a \textit{length penalty}~\cite{team2025kimi15} to prevent OOM, but these approaches also restrict the model's performance potential.

\begin{figure*}[tb]
  \setlength{\belowcaptionskip}{-5pt}
  \centering
  \includegraphics[width=0.85\linewidth]{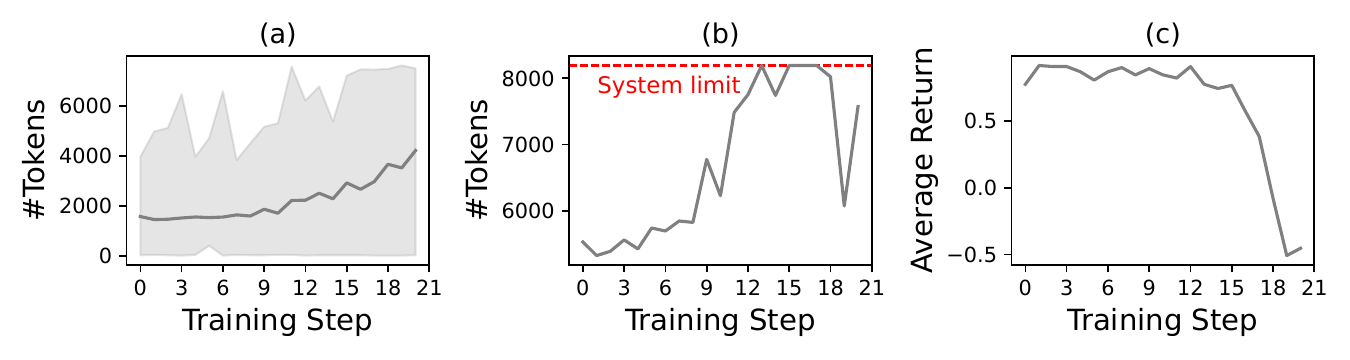}
  \caption{Training a 4B-parameter LLM on the Tic-Tac-Toe task: (a) turn-level context length steadily increases; (b) episode-level context length quickly reaches the system limit;  and (c) the model performance collapses due to context truncation.}
  \label{fig:agent_seq_length}
  \vspace{1em}
\end{figure*}

\begin{figure}[tb]
  \setlength{\belowcaptionskip}{-1em}
  \centering
  \includegraphics[width=0.95\linewidth]{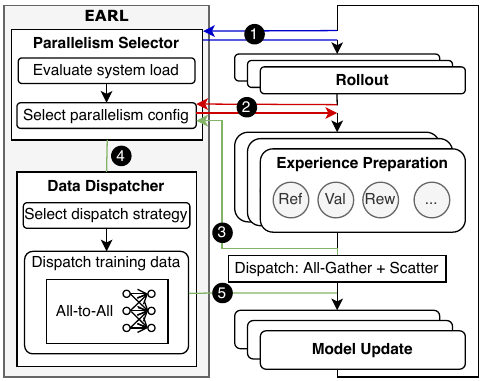}
  \caption{System design of~\ours.}
  \label{fig:arch}
\end{figure}

We observe a similar phenomenon in \emph{our industrial practice} (\F\ref{fig:agent_seq_length}): a 4B-parameter LLM is trained in a Tic-Tac-Toe environment with a maximum context length of $8{,}192$ (due to GPU memory constraints), and each episode consists of approximately three turns.
Even early in training (\F\ref{fig:agent_seq_length}a), the average single-turn response length increases steadily. \footnote{Turn-level context length refers to the token length within a single agent–environment interaction round, while episode-level context length refers to the cumulative number of tokens across an entire episode.}
By step~13 (\F\ref{fig:agent_seq_length}b), the episode-level context length reaches the system limit, causing truncated reasoning and introducing ``low-quality'' data into the rollouts. The degradation leads to a sharp drop in average return and ultimately collapses learning after step~15 (\F\ref{fig:agent_seq_length}c). 

\begin{table*}[t]
\vspace{1em}
  \centering
  \caption{Intermediate Data Batch Size Under Different Context Lengths on a 1k-GPU Cluster.}
  \begin{tabular}{c|cccccc}
    \hline
    \textbf{Context Length}                  & 1,024   & 2,048   & 4,096   & 8,192    & 16,384   & 32,768   \\
    \hline
    \textbf{Estimated Size (MiB)} & 15,625 & 31,250 & 62,500 & 125,000 & 250,000 & 500,000 \\
    \hline
  \end{tabular}
  \label{table:seq_len_mem}
  \vspace{1em}
\end{table*}

Long contexts also hinder scalability by generating massive volumes of intermediate data that must be exchanged across nodes, creating substantial communication overhead.
These intermediate batches consist of tensors required to compute training signals, including tokens, log probabilities, rewards, returns, and other auxiliary tensors.
The estimated sizes of such batches are reported in~\autoref{table:seq_len_mem}.
At the 1K-GPU scale, the aggregated data volume grows linearly with context length, reaching up to 500\unit{GB} at 32K tokens. 

\emph{In our industrial practice}, we have observed this significant \textit{data dispatch bottleneck}, exacerbated by increasing context length when scaling training to $1{,}024$ GPUs.
For instance, while training a model with over 200B parameters at context length 32K using the VeRL framework~\cite{sheng2025hybridflow}, the data volume approached 1\unit{TB} due to additional implementation overhead.
This amount of data required more than 20 minutes for transmission (under a 25 Gbps peak bandwidth), occupying over 25\% of the total iteration time and severely degrading training throughput.
The bottleneck is further aggravated by VeRL’s \textit{single-controller} architecture, in which a centralized process coordinates data exchange across different stages, forcing all intermediate data to be aggregated on a single node before redistribution.

These challenges reveal a fundamental challenge in scaling agentic RL: longer contexts boost capability but also strain memory and communication.
Existing safeguards, such as \textit{hard length limits}, mitigate resource pressure but also cap performance ceiling.
This motivates the design of~\ours, which tackles the context length explosion issue and data dispatching bottleneck, for stable and efficient large-scale training.
\vspace{-2em}
\section{\ours\ Design}
\label{sec:scaling}

We aim to scale agentic RL training to support exploding context lengths arising from response length growth and intensified multi-turn interactions, while simultaneously scaling training to thousands of GPUs.
To this end, we design ~\ours, a scalable agentic RL system with two key extensions: the \textit{Parallelism Selector} for dynamic parallelism configuration and the \textit{Data Dispatcher} for efficient inter-stage data dispatching. \looseness=-1

\F\ref{fig:arch} illustrates the design of ~\ours, highlighting the integration of these components into a standard RL training loop.
Before the Rollout stage (step~\myc{1}) and the Experience Preparation stage (step~\myc{2}), the \textit{Parallelism Selector} determines each model's parallelism configuration by evaluating the current system load and the maximum context length.
In steps~\myc{3}, \myc{4}, and \myc{5}, the \textit{Data Dispatcher} selects a layout-aware dispatch strategy using the selected parallelism and data layout of the experience preparation stage.
Once dispatch is complete, all models proceed with their respective training updates. 
We describe each component in detail below:
\looseness=-1
\tinyskip

\mypar{Parallelism Selector}
\ours applies dynamic parallelism in both the Rollout stage and experience preparation stage, configuring the policy model in the former, and the reference, value, and reward models in the latter.
The parallelism configuration is dynamically adjusted based on the current system load and the context length.
Specifically, at the start of the training process, \ours measures the throughput under various parallelism configurations and context lengths, then maintains the optimal configuration for each context length range for later use.
During training, ~\ours monitors the averaged context length generated by the model. 
When the averaged context length falls into a new context range, \ours switches to the corresponding parallelism configuration before the next Rollout stage.

\tinyskip
\mypar{Data Dispatcher}
\ours uses a data dispatch logic that is adaptive to the current data distribution layout and parallelism configuration. 
During the experience preparation stage, intermediate training batches, including tokens, log-probabilities, rewards, returns, and other tensors, must be transferred across all workers, which is a critical bottleneck with the centralized \texttt{gather-and-dispatch} mechanism in the single-controller architecture.
We introduce a parallelism- and layout-aware dispatch mechanism that sends data directly to the target workers from their computation origins, to eliminate the centralized aggregation.
Specifically, we replace the \texttt{all-gather-and-scatter} dispatch logic with an \texttt{all-to-all} operation, thereby reducing both data movement volume and synchronization overhead.
\section{Evaluation}

We evaluate the components of~\ours: (i) \textit{Parallelism Selector}~(\S\ref{sec:exp:dynamic_parallelism}) and (ii) \textit{Data Dispatcher}~(\S\ref{sec:exp:data_dispatch}) in scenarios where the context length increases during agentic RL training.

\subsection{Experiment Setup}
Our experiments have the following setup:

\tinyskip
\mypar{Testbed}
We have deployed~\ours on a cluster of 16 machines, each being equipped with 8$\times$ NVIDIA H100-80\unit{GB} GPUs (128 GPUs in total).
Intra-node GPU connection uses NVLink, while inter-node connection leverages InfiniBand with 200 Gbps.
Each node has 112 CPU cores and 1.8\unit{TB} RAM.
The software stack includes CUDA 12.4, PyTorch 2.6.0, Ray 2.46.0, and vLLM 0.8.4.
The entire execution environment is containerized, built from the NGC Docker image\footnote{nvcr.io/nvidia/pytorch:24.05-py3} with corresponding software updated to the specified versions.

\tinyskip

\mypar{Models and Training Environments}
We train Qwen2.5-72B-Instruct~\cite{qwen2.5} in an agentic setting within the Connect Four\footnote{https://en.wikipedia.org/wiki/Connect\_Four} environment.
The training begins with a tensor parallelism degree of 4, and the initial maximum context length is set to $8{,}192$.
We employ a customized agentic RL algorithm, which utilizes REINFORCE~\cite{hu2025reinforceplusplus} as the advantage estimator.
\looseness=-1

\tinyskip

\mypar{Implementation}
We have built~\ours on top of ROLL~\cite{wang_reinforcement_roll_2025}, an open-source framework for agentic RL training.
The agentic environment, Connect-Four, is implemented with open-spiel~\cite{LanctotEtAl2019OpenSpiel} and integrated into ROLL. 
The \textit{Parallelism Selector} is activated before the Rollout stage in each training step.
We optimize the data dispatch logic between the Experience Preparation stage and the Model Update stage to avoid the aggregation behavior in the single-controller architecture.

\tinyskip
\mypar{Metrics}
We evaluate the performance of the \textit{Parallelism Selector} by measuring the \textit{relative throughput speedup} of \emph{tokens-per-GPU-per-second}, which is denoted as $\text{TGS}$.
Specifically, the relative speedup of switching from $TP=a$ to $TP=b$ is:
\begin{equation}
\text{Speedup}_{\%}(a, b) = \frac{\text{TGS}(b) - \text{TGS}(a)}{\text{TGS}(a)} \times 100
\label{eq:speedup}
\end{equation}
\noindent where a positive value indicates that $TP=b$ achieves higher throughput than $TP=a$.

\subsection{Dynamic Parallelism in Rollout stage}
\label{sec:exp:dynamic_parallelism}

\begin{figure}[!t]
  \setlength{\belowcaptionskip}{-1em}
  \centering
  \includegraphics[width=0.9\linewidth]{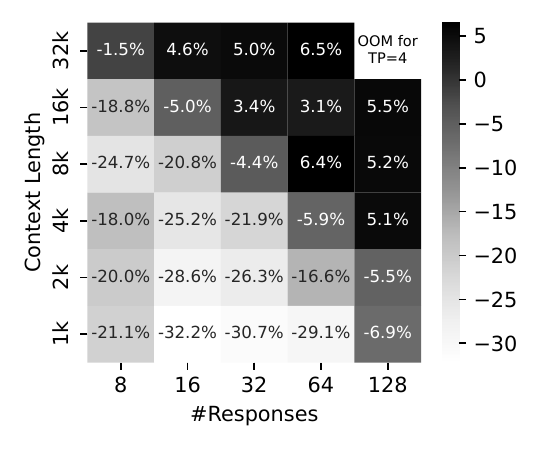}
  \caption{Relative throughput speedup from $TP=4$ to $TP=8$ across different context lengths and response counts, computed using Equation~\ref{eq:speedup}. Positive values indicate TP8 outperforms TP4; negative values indicate TP4 outperforms TP8.}
  \label{fig:perf_gap}
\end{figure}

As shown in \F\ref{fig:perf_gap}, we report $\text{Speedup}_{\%}(4, 8)$, the relative throughput improvement in the decoding phase of the Rollout stage, when switching the tensor parallelism degree from $TP=4$ to $TP=8$.
The results demonstrate the effectiveness of adapting the parallelism configuration to changes in the increasing context length during training.
In practice, the number of responses for the Rollout stage is typically fixed, while both response length and context length increase as the multi-turn training progresses.
In the case of $\#responses=32$, our approach maintains the performance advantage of $TP=4$ (31\% higher throughput) when the context length is small.
When the context length reaches 16K and 32K, ~\ours switches to $TP=8$, which yields $5\%$ improvement.
In the most extreme case, with 128 responses and a 32K context length, TP=4 encounters out-of-memory (OOM) failures, whereas switching to $TP=8$ maintains system stability and prevents crashes.

\subsection{Optimizing Data Dispatching Between Stages}
\label{sec:exp:data_dispatch}

\begin{figure}[!t]
  \setlength{\belowcaptionskip}{-1em}
  \centering
  \includegraphics[width=0.9\linewidth]{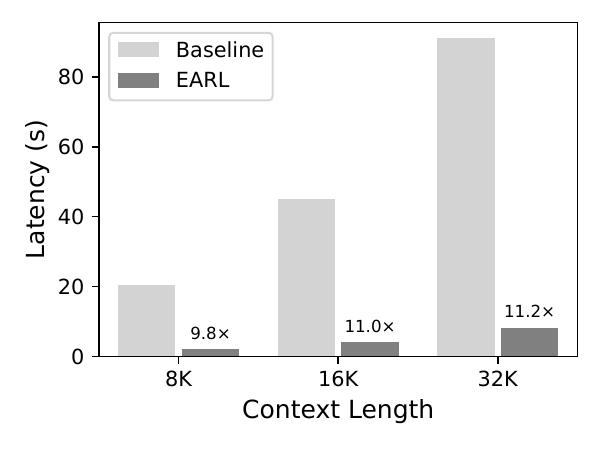}
  \caption{Data dispatch latency of baseline and~\ours under different context lengths. Numbers above the bars indicate the relative latency reduction of~\ours compared to the baseline.}
  \label{fig:data_dispatch_latency}
\end{figure}

We optimize the data dispatch logic for transferring \textit{log-probability} tensors from the reference model to the training workers, since these tensors are not required for aggregation in advantage estimation.
The intermediate data sizes are 46 MiB, 93 MiB, and 187 MiB per independent worker.
As shown in~\F\ref{fig:data_dispatch_latency}, the data dispatcher consistently achieves better performance across different context lengths.
At a context length of $8$K, the optimization reduces transmission time by $9.7\times$, and when the context length reaches $32$K, it yields up to $11.2\times$ reduction in latency.
The current prototype employs TCP over Ethernet, identical to the baseline transport, and we expect further gains with RDMA-based communication. \looseness=-1

\section{Related Work}
Efficient large-scale agentic systems are an area of active research. However, existing agentic RL systems do not optimize for the dynamic and increasing nature of context length during training and rollouts. Instead, they often rely on general inference techniques for handling long context.  
VeRL~\cite{sheng2025hybridflow}, SkyRL~\cite{griggs_evolving_skyrl_2025}, and ROLL~\cite{wang_reinforcement_roll_2025} incorporate tensor parallelism~\cite{megatron-lm} and sequence parallelism~\cite{jacobs_deepspeed_ulysses_2023, korthikanti_reducing_2022} to enable long context training. Slime~\cite{zhu_slime_2025} handles long-context during rollouts by using SGLang's chunked prefill technique~\cite{zheng_sglang_2024}. \ours complements these systems by introducing dynamic parallelism that adapts to the context length in the Rollout stage and optimizing the data dispatch logic to improve efficiency at scale.
\looseness=-1

Other approaches implicitly apply length penalties in training to constrain context growth~\cite{team2025kimi15,xiang_just_2025,aggarwal_l1_2025,golubev_training_2025}. Some works, such as SkyWork-OR1~\cite{he_skywork_2025} and DeepCoder~\cite{luo_deepcoder_2025}, progressively increase the context length across training stages to enable effective rollouts at shorter context lengths. Our work, \ours, is orthogonal to both strategies and focuses on system-level optimizations that can be utilized with any training-time technique to scale agentic RL effectively under long-context regimes.

\
\section{Limitations and Future Work}

\ours presents an initial prototype for building efficient agentic RL systems for LLMs, with a focus on addressing the challenge of context length explosion in agentic RL.
For dynamic parallelism, we have so far optimized only the \textit{Rollout} stage, without extending the optimization to the training stage. 
The Rollout stage only performs inference, which differs significantly in workload from training. 
Achieving joint optimization with the training stage requires a more comprehensive design, but we expect this direction to yield substantial performance gains.

On the other hand, in data movement, the data dispatch logic optimization focuses on tensors with minimal inter-stage dependencies (i.e., \textit{log-probabilities} are not required for advantage estimation). 
However, our approach can be applied to other tensors, such as \textit{rewards} and \textit{advantages}. 
In the current system, \textit{rewards} and \textit{returns} are aggregated for \textit{advantage} estimation. 
We will improve this process in a distributed manner to alleviate communication bottlenecks under exploding context lengths, and to better leverage \texttt{all-to-all} communication patterns for improved efficiency.

Other future directions include designing fully asynchronous RL systems for more flexible scheduling, integrating replay buffers into off-policy training to enhance data dispatch efficiency, and extending our methods to a broader class of algorithms. 
We believe these insights and advances will guide the development of more efficient and robust agentic RL systems for LLMs.

\section{Conclusion}
We address the context length explosion issue in scaling agentic RL systems and design a framework, \ours, with two core components: a \textit{Parallelism Selector} for dynamic parallelism configuration and a \textit{Data Dispatcher} for parallelism- and layout-aware data distribution, both yielding measurable performance and stability gains in large-scale training.

\section{Acknowledgments}
This work is funded by the Research Council of Finland (grant number 362729), Business Finland (grant number\\ 169/31/2024), and the Finnish Doctoral Program Network in Artificial Intelligence (AI-DOC).

\bibliographystyle{ACM-Reference-Format}
\bibliography{main}


\begin{thebibliography}{30}


\ifx \showCODEN    \undefined \def \showCODEN     #1{\unskip}     \fi
\ifx \showISBNx    \undefined \def \showISBNx     #1{\unskip}     \fi
\ifx \showISBNxiii \undefined \def \showISBNxiii  #1{\unskip}     \fi
\ifx \showISSN     \undefined \def \showISSN      #1{\unskip}     \fi
\ifx \showLCCN     \undefined \def \showLCCN      #1{\unskip}     \fi
\ifx \shownote     \undefined \def \shownote      #1{#1}          \fi
\ifx \showarticletitle \undefined \def \showarticletitle #1{#1}   \fi
\ifx \showURL      \undefined \def \showURL       {\relax}        \fi
\providecommand\bibfield[2]{#2}
\providecommand\bibinfo[2]{#2}
\providecommand\natexlab[1]{#1}
\providecommand\showeprint[2][]{arXiv:#2}

\bibitem[Aggarwal and Welleck(2025)]%
        {aggarwal_l1_2025}
\bibfield{author}{\bibinfo{person}{Pranjal Aggarwal} {and} \bibinfo{person}{Sean Welleck}.} \bibinfo{year}{2025}\natexlab{}.
\newblock \bibinfo{title}{L1: {Controlling} {How} {Long} {A} {Reasoning} {Model} {Thinks} {With} {Reinforcement} {Learning}}.
\newblock
\href{https://doi.org/10.48550/arXiv.2503.04697}{doi:\nolinkurl{10.48550/arXiv.2503.04697}}
\newblock
\shownote{arXiv:2503.04697 [cs] version: 1}.


\bibitem[Christiano et~al\mbox{.}(2017)]%
        {christiano2017deep}
\bibfield{author}{\bibinfo{person}{Paul~F Christiano}, \bibinfo{person}{Jan Leike}, \bibinfo{person}{Tom Brown}, \bibinfo{person}{Miljan Martic}, \bibinfo{person}{Shane Legg}, {and} \bibinfo{person}{Dario Amodei}.} \bibinfo{year}{2017}\natexlab{}.
\newblock \showarticletitle{Deep reinforcement learning from human preferences}.
\newblock \bibinfo{journal}{\emph{Advances in neural information processing systems}}  \bibinfo{volume}{30} (\bibinfo{year}{2017}).
\newblock


\bibitem[DeepMind(2025)]%
        {gemini_deep_research}
\bibfield{author}{\bibinfo{person}{Google DeepMind}.} \bibinfo{year}{2025}\natexlab{}.
\newblock \bibinfo{title}{{G}emini {D}eep {R}esearch — your personal research assistant --- gemini.google}.
\newblock \bibinfo{howpublished}{\url{https://gemini.google/overview/deep-research/}}.
\newblock


\bibitem[Dong et~al\mbox{.}(2025)]%
        {dong2025reinforcement}
\bibfield{author}{\bibinfo{person}{Qingxiu Dong}, \bibinfo{person}{Li Dong}, \bibinfo{person}{Yao Tang}, \bibinfo{person}{Tianzhu Ye}, \bibinfo{person}{Yutao Sun}, \bibinfo{person}{Zhifang Sui}, {and} \bibinfo{person}{Furu Wei}.} \bibinfo{year}{2025}\natexlab{}.
\newblock \showarticletitle{Reinforcement Pre-Training}.
\newblock \bibinfo{journal}{\emph{arXiv preprint arXiv:2506.08007}} (\bibinfo{year}{2025}).
\newblock


\bibitem[Golubev et~al\mbox{.}(2025)]%
        {golubev_training_2025}
\bibfield{author}{\bibinfo{person}{Alexander Golubev}, \bibinfo{person}{Maria Trofimova}, \bibinfo{person}{Sergei Polezhaev}, \bibinfo{person}{Ibragim Badertdinov}, \bibinfo{person}{Maksim Nekrashevich}, \bibinfo{person}{Anton Shevtsov}, \bibinfo{person}{Simon Karasik}, \bibinfo{person}{Sergey Abramov}, \bibinfo{person}{Andrei Andriushchenko}, \bibinfo{person}{Filipp Fisin}, \bibinfo{person}{Sergei Skvortsov}, {and} \bibinfo{person}{Boris Yangel}.} \bibinfo{year}{2025}\natexlab{}.
\newblock \bibinfo{title}{Training {Long}-{Context}, {Multi}-{Turn} {Software} {Engineering} {Agents} with {Reinforcement} {Learning}}.
\newblock
\href{https://doi.org/10.48550/arXiv.2508.03501}{doi:\nolinkurl{10.48550/arXiv.2508.03501}}
\newblock
\shownote{arXiv:2508.03501 [cs] version: 1}.


\bibitem[Griggs et~al\mbox{.}(2025)]%
        {griggs_evolving_skyrl_2025}
\bibfield{author}{\bibinfo{person}{Tyler Griggs}, \bibinfo{person}{Sumanth Hegde}, \bibinfo{person}{Eric Tang}, \bibinfo{person}{Shu Liu}, \bibinfo{person}{Shiyi Cao}, \bibinfo{person}{Dacheng Li}, \bibinfo{person}{Charlie Ruan}, \bibinfo{person}{Philipp Moritz}, \bibinfo{person}{Kourosh Hakhamaneshi}, \bibinfo{person}{Richard Liaw}, \bibinfo{person}{Akshay Malik}, \bibinfo{person}{Matei Zaharia}, \bibinfo{person}{Joseph~E. Gonzalez}, {and} \bibinfo{person}{Ion Stoica}.} \bibinfo{year}{2025}\natexlab{}.
\newblock \bibinfo{title}{Evolving {SkyRL} into a {Highly}-{Modular} {RL} {Framework}}.
\newblock


\bibitem[Guo et~al\mbox{.}(2025)]%
        {guo2025deepseekr1}
\bibfield{author}{\bibinfo{person}{Daya Guo}, \bibinfo{person}{Dejian Yang}, \bibinfo{person}{Haowei Zhang}, \bibinfo{person}{Junxiao Song}, \bibinfo{person}{Ruoyu Zhang}, \bibinfo{person}{Runxin Xu}, \bibinfo{person}{Qihao Zhu}, \bibinfo{person}{Shirong Ma}, \bibinfo{person}{Peiyi Wang}, \bibinfo{person}{Xiao Bi}, {et~al\mbox{.}}} \bibinfo{year}{2025}\natexlab{}.
\newblock \showarticletitle{Deepseek-r1: Incentivizing reasoning capability in llms via reinforcement learning}.
\newblock \bibinfo{journal}{\emph{arXiv preprint arXiv:2501.12948}} (\bibinfo{year}{2025}).
\newblock


\bibitem[He et~al\mbox{.}(2025)]%
        {he_skywork_2025}
\bibfield{author}{\bibinfo{person}{Jujie He}, \bibinfo{person}{Jiacai Liu}, \bibinfo{person}{Chris~Yuhao Liu}, \bibinfo{person}{Rui Yan}, \bibinfo{person}{Chaojie Wang}, \bibinfo{person}{Peng Cheng}, \bibinfo{person}{Xiaoyu Zhang}, \bibinfo{person}{Fuxiang Zhang}, \bibinfo{person}{Jiacheng Xu}, \bibinfo{person}{Wei Shen}, \bibinfo{person}{Siyuan Li}, \bibinfo{person}{Liang Zeng}, \bibinfo{person}{Tianwen Wei}, \bibinfo{person}{Cheng Cheng}, \bibinfo{person}{Bo An}, \bibinfo{person}{Yang Liu}, {and} \bibinfo{person}{Yahui Zhou}.} \bibinfo{year}{2025}\natexlab{}.
\newblock \bibinfo{title}{Skywork {Open} {Reasoner} 1 {Technical} {Report}}.
\newblock
\href{https://doi.org/10.48550/arXiv.2505.22312}{doi:\nolinkurl{10.48550/arXiv.2505.22312}}
\newblock
\shownote{arXiv:2505.22312 [cs]}.


\bibitem[Hu et~al\mbox{.}(2025)]%
        {hu2025reinforceplusplus}
\bibfield{author}{\bibinfo{person}{Jian Hu}, \bibinfo{person}{Jason~Klein Liu}, \bibinfo{person}{Haotian Xu}, {and} \bibinfo{person}{Wei Shen}.} \bibinfo{year}{2025}\natexlab{}.
\newblock \showarticletitle{Reinforce++: An efficient rlhf algorithm with robustness to both prompt and reward models}.
\newblock \bibinfo{journal}{\emph{arXiv preprint arXiv:2501.03262}} (\bibinfo{year}{2025}).
\newblock


\bibitem[Jacobs et~al\mbox{.}(2023)]%
        {jacobs_deepspeed_ulysses_2023}
\bibfield{author}{\bibinfo{person}{Sam~Ade Jacobs}, \bibinfo{person}{Masahiro Tanaka}, \bibinfo{person}{Chengming Zhang}, \bibinfo{person}{Minjia Zhang}, \bibinfo{person}{Shuaiwen~Leon Song}, \bibinfo{person}{Samyam Rajbhandari}, {and} \bibinfo{person}{Yuxiong He}.} \bibinfo{year}{2023}\natexlab{}.
\newblock \bibinfo{title}{{DeepSpeed} {Ulysses}: {System} {Optimizations} for {Enabling} {Training} of {Extreme} {Long} {Sequence} {Transformer} {Models}}.
\newblock
\href{https://doi.org/10.48550/arXiv.2309.14509}{doi:\nolinkurl{10.48550/arXiv.2309.14509}}
\newblock
\shownote{arXiv:2309.14509 [cs]}.


\bibitem[Jern et~al\mbox{.}(2025)]%
        {jern_agent-q_2025}
\bibfield{author}{\bibinfo{person}{Linus Jern}, \bibinfo{person}{Valter Uotila}, \bibinfo{person}{Cong Yu}, {and} \bibinfo{person}{Bo Zhao}.} \bibinfo{year}{2025}\natexlab{}.
\newblock \bibinfo{title}{Agent-{Q}: {Fine}-{Tuning} {Large} {Language} {Models} for {Quantum} {Circuit} {Generation} and {Optimization}}.
\newblock
\href{https://doi.org/10.48550/arXiv.2504.11109}{doi:\nolinkurl{10.48550/arXiv.2504.11109}}
\newblock
\shownote{arXiv:2504.11109 [quant-ph] version: 2}.


\bibitem[Korthikanti et~al\mbox{.}(2022)]%
        {korthikanti_reducing_2022}
\bibfield{author}{\bibinfo{person}{Vijay Korthikanti}, \bibinfo{person}{Jared Casper}, \bibinfo{person}{Sangkug Lym}, \bibinfo{person}{Lawrence McAfee}, \bibinfo{person}{Michael Andersch}, \bibinfo{person}{Mohammad Shoeybi}, {and} \bibinfo{person}{Bryan Catanzaro}.} \bibinfo{year}{2022}\natexlab{}.
\newblock \bibinfo{title}{Reducing {Activation} {Recomputation} in {Large} {Transformer} {Models}}.
\newblock
\href{https://doi.org/10.48550/arXiv.2205.05198}{doi:\nolinkurl{10.48550/arXiv.2205.05198}}
\newblock
\shownote{arXiv:2205.05198 [cs]}.


\bibitem[Lanctot et~al\mbox{.}(2019)]%
        {LanctotEtAl2019OpenSpiel}
\bibfield{author}{\bibinfo{person}{Marc Lanctot}, \bibinfo{person}{Edward Lockhart}, \bibinfo{person}{Jean-Baptiste Lespiau}, \bibinfo{person}{Vinicius Zambaldi}, \bibinfo{person}{Satyaki Upadhyay}, \bibinfo{person}{Julien P\'{e}rolat}, \bibinfo{person}{Sriram Srinivasan}, \bibinfo{person}{Finbarr Timbers}, \bibinfo{person}{Karl Tuyls}, \bibinfo{person}{Shayegan Omidshafiei}, \bibinfo{person}{Daniel Hennes}, \bibinfo{person}{Dustin Morrill}, \bibinfo{person}{Paul Muller}, \bibinfo{person}{Timo Ewalds}, \bibinfo{person}{Ryan Faulkner}, \bibinfo{person}{J\'{a}nos Kram\'{a}r}, \bibinfo{person}{Bart~De Vylder}, \bibinfo{person}{Brennan Saeta}, \bibinfo{person}{James Bradbury}, \bibinfo{person}{David Ding}, \bibinfo{person}{Sebastian Borgeaud}, \bibinfo{person}{Matthew Lai}, \bibinfo{person}{Julian Schrittwieser}, \bibinfo{person}{Thomas Anthony}, \bibinfo{person}{Edward Hughes}, \bibinfo{person}{Ivo Danihelka}, {and} \bibinfo{person}{Jonah Ryan-Davis}.} \bibinfo{year}{2019}\natexlab{}.
\newblock \showarticletitle{{OpenSpiel}: A Framework for Reinforcement Learning in Games}.
\newblock \bibinfo{journal}{\emph{CoRR}}  \bibinfo{volume}{abs/1908.09453} (\bibinfo{year}{2019}).
\newblock
\showeprint[arxiv]{1908.09453}~[cs.LG]
\urldef\tempurl%
\url{http://arxiv.org/abs/1908.09453}
\showURL{%
\tempurl}


\bibitem[Llama~Team(2024)]%
        {dubey2024llama}
\bibfield{author}{\bibinfo{person}{AI~@~Meta Llama~Team}.} \bibinfo{year}{2024}\natexlab{}.
\newblock \showarticletitle{The llama 3 herd of models}.
\newblock \bibinfo{journal}{\emph{arXiv e-prints}} (\bibinfo{year}{2024}), \bibinfo{pages}{arXiv--2407}.
\newblock


\bibitem[Luo et~al\mbox{.}(2025)]%
        {luo_deepcoder_2025}
\bibfield{author}{\bibinfo{person}{Michael Luo}, \bibinfo{person}{Sijun Tan}, \bibinfo{person}{Roy Huang}, \bibinfo{person}{Ameen Patel}, \bibinfo{person}{Alpay Ariyak}, \bibinfo{person}{Qingyang Wu}, \bibinfo{person}{Xiaoxiang Shi}, \bibinfo{person}{Rachel Xin}, \bibinfo{person}{Colin Cai}, \bibinfo{person}{Maurice Weber}, \bibinfo{person}{Ce Zhang}, \bibinfo{person}{Li~Erran Li}, \bibinfo{person}{Raluca~Ada Popa}, {and} \bibinfo{person}{Ion Stoica}.} \bibinfo{year}{2025}\natexlab{}.
\newblock \bibinfo{title}{{DeepCoder}: {A} {Fully} {Open}‑{Source} {14B} {Coder} at {O3}‑mini {Level}}.
\newblock
\urldef\tempurl%
\url{https://www.together.ai/blog/deepcoder}
\showURL{%
\tempurl}


\bibitem[{OpenAI}(2025)]%
        {openai_deep_research}
\bibfield{author}{\bibinfo{person}{{OpenAI}}.} \bibinfo{year}{2025}\natexlab{}.
\newblock \bibinfo{title}{Introducing Deep Research}.
\newblock \bibinfo{howpublished}{\url{https://openai.com/index/introducing-deep-research}}.
\newblock


\bibitem[OpenAI(2025)]%
        {openai_gpt5_2025}
\bibfield{author}{\bibinfo{person}{OpenAI}.} \bibinfo{year}{2025}\natexlab{}.
\newblock \bibinfo{title}{Introducing GPT-5}.
\newblock
\urldef\tempurl%
\url{https://openai.com/index/introducing-gpt-5}
\showURL{%
\tempurl}


\bibitem[Ouyang et~al\mbox{.}(2022)]%
        {ouyang_training_2022}
\bibfield{author}{\bibinfo{person}{Long Ouyang}, \bibinfo{person}{Jeff Wu}, \bibinfo{person}{Xu Jiang}, \bibinfo{person}{Diogo Almeida}, \bibinfo{person}{Carroll~L. Wainwright}, \bibinfo{person}{Pamela Mishkin}, \bibinfo{person}{Chong Zhang}, \bibinfo{person}{Sandhini Agarwal}, \bibinfo{person}{Katarina Slama}, \bibinfo{person}{Alex Ray}, \bibinfo{person}{John Schulman}, \bibinfo{person}{Jacob Hilton}, \bibinfo{person}{Fraser Kelton}, \bibinfo{person}{Luke Miller}, \bibinfo{person}{Maddie Simens}, \bibinfo{person}{Amanda Askell}, \bibinfo{person}{Peter Welinder}, \bibinfo{person}{Paul Christiano}, \bibinfo{person}{Jan Leike}, {and} \bibinfo{person}{Ryan Lowe}.} \bibinfo{year}{2022}\natexlab{}.
\newblock \bibinfo{title}{Training language models to follow instructions with human feedback}.
\newblock
\href{https://doi.org/10.48550/arXiv.2203.02155}{doi:\nolinkurl{10.48550/arXiv.2203.02155}}
\newblock
\shownote{arXiv:2203.02155 [cs]}.


\bibitem[Sheng et~al\mbox{.}(2025)]%
        {sheng2025hybridflow}
\bibfield{author}{\bibinfo{person}{Guangming Sheng}, \bibinfo{person}{Chi Zhang}, \bibinfo{person}{Zilingfeng Ye}, \bibinfo{person}{Xibin Wu}, \bibinfo{person}{Wang Zhang}, \bibinfo{person}{Ru Zhang}, \bibinfo{person}{Yanghua Peng}, \bibinfo{person}{Haibin Lin}, {and} \bibinfo{person}{Chuan Wu}.} \bibinfo{year}{2025}\natexlab{}.
\newblock \showarticletitle{Hybridflow: A flexible and efficient rlhf framework}. In \bibinfo{booktitle}{\emph{Proceedings of the Twentieth European Conference on Computer Systems}}. \bibinfo{pages}{1279--1297}.
\newblock


\bibitem[Shoeybi et~al\mbox{.}(2019)]%
        {megatron-lm}
\bibfield{author}{\bibinfo{person}{Mohammad Shoeybi}, \bibinfo{person}{Mostofa Patwary}, \bibinfo{person}{Raul Puri}, \bibinfo{person}{Patrick LeGresley}, \bibinfo{person}{Jared Casper}, {and} \bibinfo{person}{Bryan Catanzaro}.} \bibinfo{year}{2019}\natexlab{}.
\newblock \showarticletitle{Megatron-LM: Training Multi-Billion Parameter Language Models Using Model Parallelism}.
\newblock \bibinfo{journal}{\emph{arXiv preprint arXiv:1909.08053}} (\bibinfo{year}{2019}).
\newblock


\bibitem[Tazi et~al\mbox{.}(2025)]%
        {ultrascale_playbook}
\bibfield{author}{\bibinfo{person}{Nouamane Tazi}, \bibinfo{person}{Ferdinand Mom}, \bibinfo{person}{Haojun Zhao}, \bibinfo{person}{Phuc Nguyen}, \bibinfo{person}{Mohamed Mekkouri}, \bibinfo{person}{Leandro Werra}, {and} \bibinfo{person}{Thomas Wolf}.} \bibinfo{year}{2025}\natexlab{}.
\newblock \bibinfo{title}{The Ultra-Scale Playbook: Training LLMs on GPU Clusters}.
\newblock
\urldef\tempurl%
\url{https://huggingface.co/spaces/nanotron/ultrascale-playbook}
\showURL{%
\tempurl}


\bibitem[Team(2025a)]%
        {team2025kimi15}
\bibfield{author}{\bibinfo{person}{Kimi Team}.} \bibinfo{year}{2025}\natexlab{a}.
\newblock \showarticletitle{Kimi k1. 5: Scaling reinforcement learning with llms}.
\newblock \bibinfo{journal}{\emph{arXiv preprint arXiv:2501.12599}} (\bibinfo{year}{2025}).
\newblock


\bibitem[Team et~al\mbox{.}(2025)]%
        {team2025kimi}
\bibfield{author}{\bibinfo{person}{Kimi Team}, \bibinfo{person}{Yifan Bai}, \bibinfo{person}{Yiping Bao}, \bibinfo{person}{Guanduo Chen}, \bibinfo{person}{Jiahao Chen}, \bibinfo{person}{Ningxin Chen}, \bibinfo{person}{Ruijue Chen}, \bibinfo{person}{Yanru Chen}, \bibinfo{person}{Yuankun Chen}, \bibinfo{person}{Yutian Chen}, {et~al\mbox{.}}} \bibinfo{year}{2025}\natexlab{}.
\newblock \showarticletitle{Kimi K2: Open Agentic Intelligence}.
\newblock \bibinfo{journal}{\emph{arXiv preprint arXiv:2507.20534}} (\bibinfo{year}{2025}).
\newblock


\bibitem[Team(2024)]%
        {qwen2.5}
\bibfield{author}{\bibinfo{person}{Qwen Team}.} \bibinfo{year}{2024}\natexlab{}.
\newblock \bibinfo{title}{Qwen2.5: A Party of Foundation Models}.
\newblock
\urldef\tempurl%
\url{https://qwenlm.github.io/blog/qwen2.5/}
\showURL{%
\tempurl}


\bibitem[Team(2025b)]%
        {yang_qwen3_2025}
\bibfield{author}{\bibinfo{person}{Qwen Team}.} \bibinfo{year}{2025}\natexlab{b}.
\newblock \bibinfo{title}{Qwen3 {Technical} {Report}}.
\newblock
\href{https://doi.org/10.48550/arXiv.2505.09388}{doi:\nolinkurl{10.48550/arXiv.2505.09388}}
\newblock
\shownote{arXiv:2505.09388 [cs]}.


\bibitem[Wang et~al\mbox{.}(2025)]%
        {wang_reinforcement_roll_2025}
\bibfield{author}{\bibinfo{person}{Weixun Wang}, \bibinfo{person}{Shaopan Xiong}, \bibinfo{person}{Gengru Chen}, \bibinfo{person}{Wei Gao}, \bibinfo{person}{Sheng Guo}, \bibinfo{person}{Yancheng He}, \bibinfo{person}{Ju Huang}, \bibinfo{person}{Jiaheng Liu}, \bibinfo{person}{Zhendong Li}, \bibinfo{person}{Xiaoyang Li}, \bibinfo{person}{Zichen Liu}, \bibinfo{person}{Haizhou Zhao}, \bibinfo{person}{Dakai An}, \bibinfo{person}{Lunxi Cao}, \bibinfo{person}{Qiyang Cao}, \bibinfo{person}{Wanxi Deng}, \bibinfo{person}{Feilei Du}, \bibinfo{person}{Yiliang Gu}, \bibinfo{person}{Jiahe Li}, \bibinfo{person}{Xiang Li}, \bibinfo{person}{Mingjie Liu}, \bibinfo{person}{Yijia Luo}, \bibinfo{person}{Zihe Liu}, \bibinfo{person}{Yadao Wang}, \bibinfo{person}{Pei Wang}, \bibinfo{person}{Tianyuan Wu}, \bibinfo{person}{Yanan Wu}, \bibinfo{person}{Yuheng Zhao}, \bibinfo{person}{Shuaibing Zhao}, \bibinfo{person}{Jin Yang}, \bibinfo{person}{Siran Yang}, \bibinfo{person}{Yingshui Tan}, \bibinfo{person}{Huimin Yi},
  \bibinfo{person}{Yuchi Xu}, \bibinfo{person}{Yujin Yuan}, \bibinfo{person}{Xingyao Zhang}, \bibinfo{person}{Lin Qu}, \bibinfo{person}{Wenbo Su}, \bibinfo{person}{Wei Wang}, \bibinfo{person}{Jiamang Wang}, {and} \bibinfo{person}{Bo Zheng}.} \bibinfo{year}{2025}\natexlab{}.
\newblock \bibinfo{title}{Reinforcement {Learning} {Optimization} for {Large}-{Scale} {Learning}: {An} {Efficient} and {User}-{Friendly} {Scaling} {Library}}.
\newblock
\href{https://doi.org/10.48550/arXiv.2506.06122}{doi:\nolinkurl{10.48550/arXiv.2506.06122}}
\newblock
\shownote{arXiv:2506.06122 [cs]}.


\bibitem[Xiang et~al\mbox{.}(2025)]%
        {xiang_just_2025}
\bibfield{author}{\bibinfo{person}{Violet Xiang}, \bibinfo{person}{Chase Blagden}, \bibinfo{person}{Rafael Rafailov}, \bibinfo{person}{Nathan Lile}, \bibinfo{person}{Sang Truong}, \bibinfo{person}{Chelsea Finn}, {and} \bibinfo{person}{Nick Haber}.} \bibinfo{year}{2025}\natexlab{}.
\newblock \bibinfo{title}{Just {Enough} {Thinking}: {Efficient} {Reasoning} with {Adaptive} {Length} {Penalties} {Reinforcement} {Learning}}.
\newblock
\href{https://doi.org/10.48550/arXiv.2506.05256}{doi:\nolinkurl{10.48550/arXiv.2506.05256}}
\newblock
\shownote{arXiv:2506.05256 [cs] version: 1}.


\bibitem[Yu et~al\mbox{.}(2025)]%
        {yu_quasar_2025}
\bibfield{author}{\bibinfo{person}{Cong Yu}, \bibinfo{person}{Valter Uotila}, \bibinfo{person}{Shilong Deng}, \bibinfo{person}{Qingyuan Wu}, \bibinfo{person}{Tuo Shi}, \bibinfo{person}{Songlin Jiang}, \bibinfo{person}{Lei You}, {and} \bibinfo{person}{Bo Zhao}.} \bibinfo{year}{2025}\natexlab{}.
\newblock \bibinfo{title}{{QUASAR}: {Quantum} {Assembly} {Code} {Generation} {Using} {Tool}-{Augmented} {LLMs} via {Agentic} {RL}}.
\newblock
\href{https://doi.org/10.48550/arXiv.2510.00967}{doi:\nolinkurl{10.48550/arXiv.2510.00967}}
\newblock
\shownote{arXiv:2510.00967 [cs]}.


\bibitem[Zheng et~al\mbox{.}(2024)]%
        {zheng_sglang_2024}
\bibfield{author}{\bibinfo{person}{Lianmin Zheng}, \bibinfo{person}{Liangsheng Yin}, \bibinfo{person}{Zhiqiang Xie}, \bibinfo{person}{Chuyue Sun}, \bibinfo{person}{Jeff Huang}, \bibinfo{person}{Cody~Hao Yu}, \bibinfo{person}{Shiyi Cao}, \bibinfo{person}{Christos Kozyrakis}, \bibinfo{person}{Ion Stoica}, \bibinfo{person}{Joseph~E. Gonzalez}, \bibinfo{person}{Clark Barrett}, {and} \bibinfo{person}{Ying Sheng}.} \bibinfo{year}{2024}\natexlab{}.
\newblock \bibinfo{title}{{SGLang}: {Efficient} {Execution} of {Structured} {Language} {Model} {Programs}}.
\newblock
\href{https://doi.org/10.48550/arXiv.2312.07104}{doi:\nolinkurl{10.48550/arXiv.2312.07104}}
\newblock
\shownote{arXiv:2312.07104 [cs]}.


\bibitem[Zhu et~al\mbox{.}(2025)]%
        {zhu_slime_2025}
\bibfield{author}{\bibinfo{person}{Zilin Zhu}, \bibinfo{person}{Chengxing Xie}, \bibinfo{person}{Xin Lv}, {and} \bibinfo{person}{slime Contributors}.} \bibinfo{year}{2025}\natexlab{}.
\newblock \bibinfo{title}{slime: {An} {LLM} post-training framework for {RL} {Scaling}}.
\newblock
\urldef\tempurl%
\url{https://github.com/THUDM/slime}
\showURL{%
\tempurl}


\end{thebibliography}

\end{document}